\documentclass[12pt, border=55mm,preprintnumbers]{article}

\usepackage{amssymb}
\usepackage{amsmath}

\usepackage{graphicx}
\usepackage{color} 
\usepackage{relsize} 
\usepackage{lmodern}

\usepackage{hyperref}
\usepackage{enumitem}

\usepackage{float}

\def\bea{\begin{eqnarray}}
\def\eea{\end{eqnarray}}
\def\bean{\begin{eqnarray*}}
\def\eean{\end{eqnarray*}}

\def\bea{\begin{eqnarray}}
\def\eea{\end{eqnarray}}
\def\bean{\begin{equation*}}
\def\eean{\end{equation*}}

\begin{document}

\thispagestyle{empty}

\noindent\
\\
\\
\\
\begin{center}
\large \bf Is There a Sign of New Physics \\
in Beryllium Transitions?\,\footnote{Invited talk given at the American Physical Society April Meeting 2017, Washington, DC, January 28, 2017;
based on the work done in collaboration with Jonathan Feng, Iftah Galon, Susan Gardner, Jordan Smolinsky, Tim Tait and Philip Tanedo \cite{Feng:2016jff,Feng:2016ysn}.}
\end{center}
\hfill
 \vspace*{1cm}
\noindent
\begin{center}
{\bf Bartosz Fornal}\\ \vspace{2mm}
{\emph{Department of Physics, University of California, San Diego \\
9500 Gilman Drive, La Jolla, CA 92093, USA}}
\vspace*{1.5cm}
\end{center}

\begin{abstract}
We discuss the current status of the anomaly in beryllium-8 nuclear transitions recently reported in the angular distribution of  internal  conversion electron-positron pairs. We present a phenomenological analysis of the signal and review the models proposed to explain it, focusing on those involving a new light protophobic vector gauge boson. We also elaborate on the prospects of verifying  the anomaly in  present and future experiments.
\end{abstract}

\newpage

\section{Why  New Physics?}\label{intro}
Our current understanding of Nature at the fundamental level is successfully and elegantly captured by the Standard Model of elementary particles. Sadly, this theory describes only five percent of the entire content of the Universe. The  overwhelming unknown constituents are dark matter and dark energy. Although not much is known about the origin of dark energy, we have quite compelling reasons to believe that dark matter is made up of new kind of particles. The experimental limits on their non-gravitational interactions with the particles of the Standard Model are severe, forcing those interactions to be very weak, if at all nonzero. The sole existence of dark matter is the key motivator for new particles searches. 

The Standard Model is based on the local symmetry ${\rm SU(3)}_c \times {\rm SU(2)}_L \times {\rm U(1)}_Y$   \cite{Glashow:1961tr,Weinberg:1967tq,Salam:1968rm,SU(3),Fritzsch:1973pi}. All fundamental  particles discovered so far come in representations of this gauge group. It seems reasonable to expect that dark matter also fits into this general picture,  carrying charges under a new gauge group, with a very small mixing with the Standard Model. The simplest realization of this idea is to postulate a new ``dark'' ${\rm U(1)}$ local symmetry, forming a hidden sector in which  the dark matter resides. In such a scenario the dark matter can communicate with the Standard Model only through a new gauge boson mediating interactions between the two sectors. If such a gauge boson has small couplings to the Standard Model particles, it could have escaped experimental detection even if it is light, on the scale of a few MeV   \cite{Alexander:2016aln,Battaglieri:2017aum}. 

\begin{figure}[t!]
  \centering
      \includegraphics[width=1\textwidth]{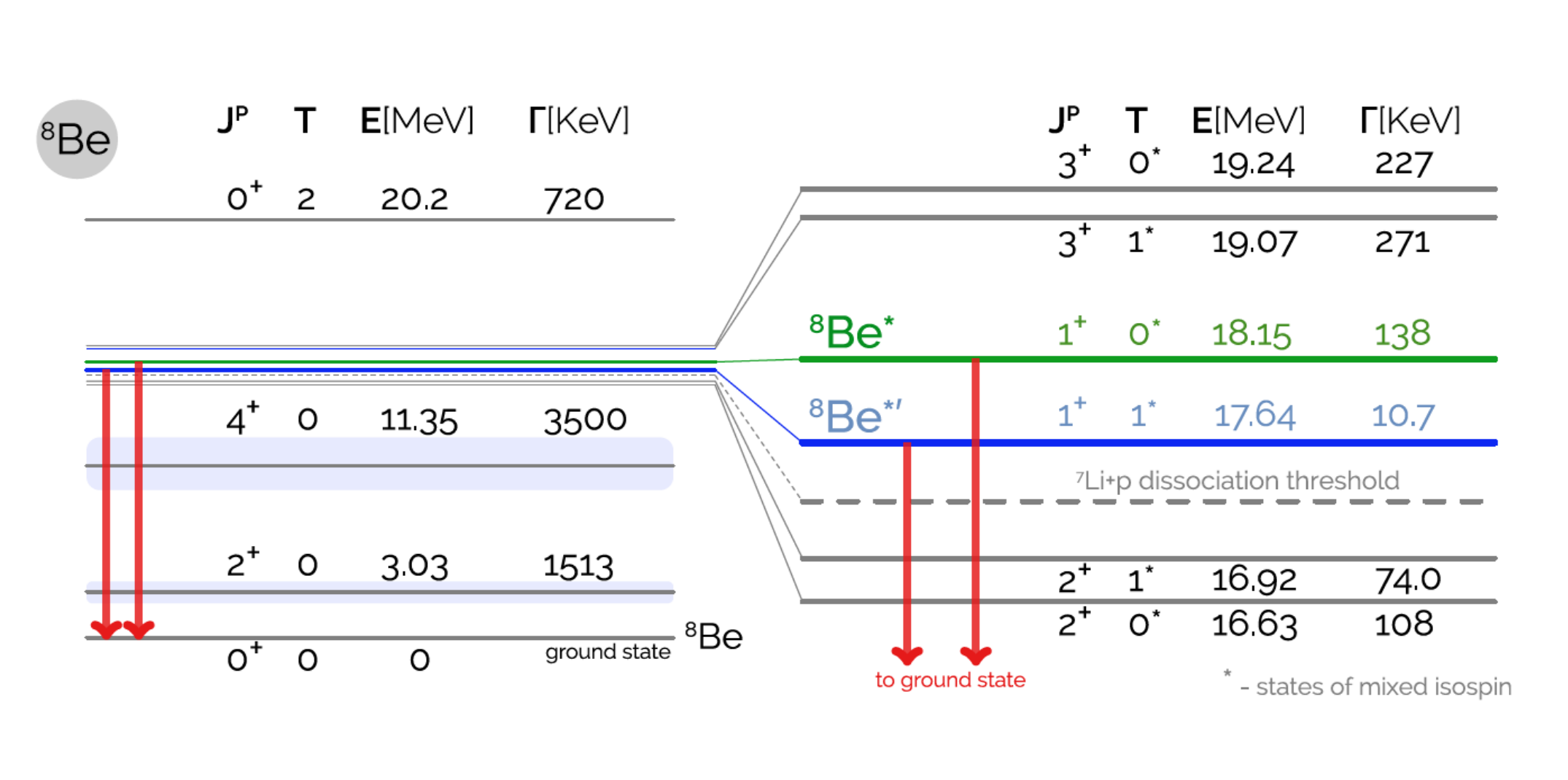}\vspace{-5mm}
  \caption{\small{Relevant $^8{\rm Be}$ nuclear energy levels and the transitions measured by the Atomki experiment. Figure adopted from   \cite{Feng:2016ysn}.}}\vspace{7mm}
  \label{fig:1}
\end{figure}

\begin{figure}[t!]
  \centering
      \includegraphics[width=0.3\textwidth]{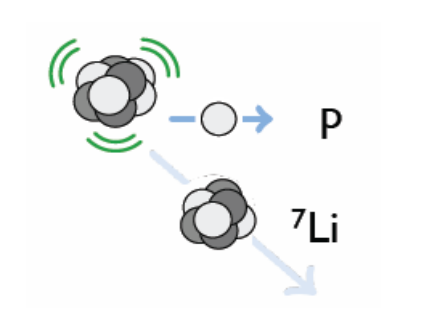}
            \includegraphics[width=0.3\textwidth]{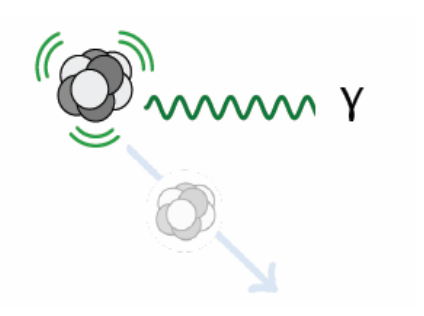}
                  \includegraphics[width=0.3\textwidth]{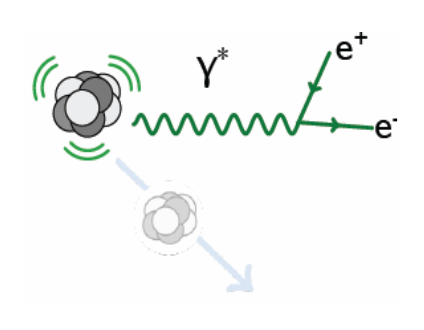}
  \caption{\small{$^8{\rm Be}(18.15)$  decay channels: hadronic, electromagnetic and through internal pair conversion. \vspace{5mm}}}
  \label{fig:234}
\end{figure}

It is quite difficult to look for light  bosons at particle colliders. Thankfully, as pointed out already a long time ago   \cite{Treiman:1978ge,Donnelly:1978ty}, nuclear transitions provide powerful probes of MeV-scale new physics, especially in systems with a large splitting between nuclear energy levels  connected via electromagnetic transitions. This has been used over the years to look for light bosons   \cite{Freedman:1984sd,Savage:1986ty,Savage:1988rg,PhysRevLett.61.1274,PhysRevLett.62.2639,deBoer1996235,deBoer:1997mr,deBoer:1999tw,deBoer:2001sjo,2006AcPPB..37..239K,deBoer:2010qb}, though without any strong evidence. However, recently an experiment looking at electron-positron internal conversion pairs in beryllium-8 ($^8{\rm Be}$) claimed a 6.8 sigma evidence for an anomaly   \cite{Krasznahorkay:2015iga}. Before describing those results, let us first discuss why the $^8{\rm Be}$ nucleus offers one of the most appealing environments for this type of search.

\section{New Physics Search in Beryllium}
\label{newphysics}
The $^8{\rm Be}$ nucleus consists of four protons and four neutrons. Its uniqueness arises from the fact that it contains narrow states with unusually high energies ($17.64 \ \rm MeV$ and $18.15 \ \rm MeV$) which decay in part to the ground state through electromagnetic transitions (see Fig.~\ref{fig:1}). These excited states of $^8{\rm Be}$ can be easily produced through $p + \!\,\!^7{\rm Li}$ reaction with high statistics, therefore creating a perfect environment to search for new MeV-scale new physics. Some of just a handful of other nuclear levels decaying via  such energetic discrete electromagnetic transitions  are: $^{10}{\rm Be}(17.8)$, $^{10}{\rm B}(18.4)$ and $^{10}{\rm B}(19.3)$   \cite{Ling:1971oik,Subotic:1978mab}. 

 \begin{figure}[t!]
  \centering
      \includegraphics[width=1\textwidth]{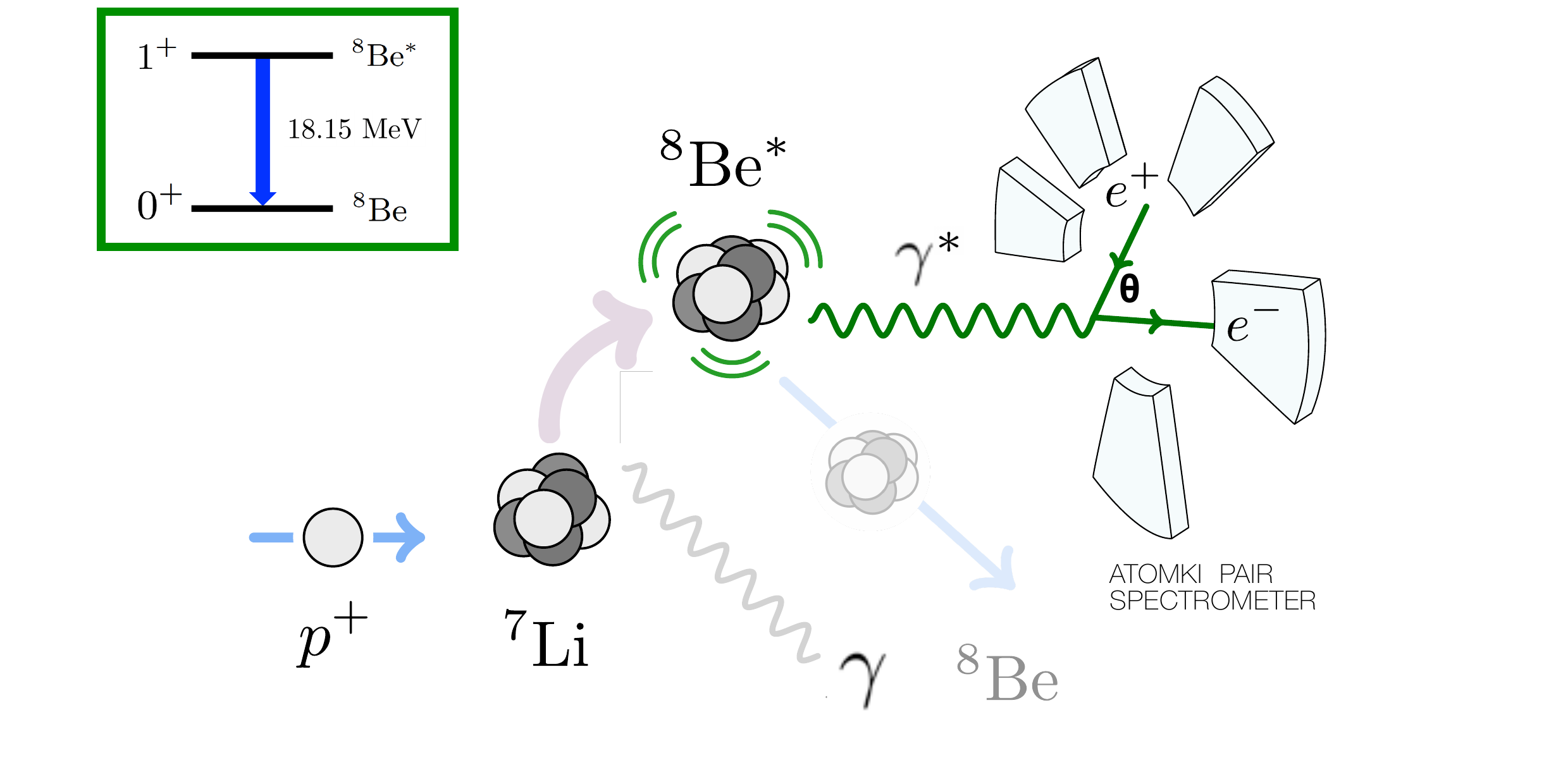}\vspace{-7mm}
  \caption{\small{A schematic visualization of the Atomki experimental setup. Figure adopted from \cite{Feng:2016ysn}.\vspace{2mm}}}
  \label{fig:5}
\end{figure}

The state which we will be particularly interested in is $^8{\rm Be}(18.15)$. Its decay  modes are shown in Fig. 2, with the absolute dominance of the  hadronic decay to $p + \,\!^7{\rm Li}$. The branching ratio for an electromagnetic transition to the ground state is  $\sim 1.5\times 10^{-5}$   \cite{Tilley:2004zz}. It is an isoscalar magnetic transition from a state with total angular momentum one ($J\!=\!1$), even parity ($P\!=\!1$) and isospin zero ($T\!=\!0$)\footnote{\label{foot1}The situation is actually more complicated since the state $^8{\rm Be}(18.15)$ is not a pure isospin state. The isospin  eigenstate is a linear combination of  $^8{\rm Be}(18.15)$ and  $^8{\rm Be}(17.64)$  \cite{Pastore:2014oda}. We will discuss this in Sec.~\ref{isospinmixing2}} to the ground state with $J\!=\!0$, $P\!=\!1$ and $T\!=\!0$. Each electromagnetic decay with energy higher than $2m_e$ is accompanied by a process involving internal pair conversion, with an electron-positron pair in the final state instead of a photon. Such a process is suppressed by $\alpha_{\rm EM}$ and in the $^8{\rm Be}(18.15)$ case its branching ratio is  $\sim 5.5\times 10^{-8}$ \cite{Rose:1949zz,Schluter1981327}. The  angular distribution of the internal conversion electron-positron pairs has been calculated a long time ago  \cite{Rose:1949zz,Horton}. It is sharply peaked at low electron-positron opening angles and then monotonically decreases in the direction of larger  angles.

\section{Beryllium Anomaly}

The Atomki experiment  \cite{Krasznahorkay:2015iga} measured precisely the angular distribution of such electron-positron pairs. A beam of protons with tunable energy was directed on a stationary target containing $^7{\rm Li}$ nuclei. For the resonant proton kinetic energy of $1.04 \ {\rm MeV}$ the reaction $p \,+ \,\!^7{\rm Li} \rightarrow\,  ^8{\rm Be}(18.15) + \gamma$ took place, followed by the $^8{\rm Be}(18.15)$ decay, in part through internal pair conversion. The resulting electron-positron pairs were recorded with high statistics and improved accuracy compared to earlier experiments  (see Fig.~\ref{fig:5}). For details on the experimental setup and the Atomki spectrometer see  \cite{Krasznahorkay:2015iga,Gulyas:2015mia}. 

\begin{figure}[t!]
  \centering
      \includegraphics[width=1\textwidth]{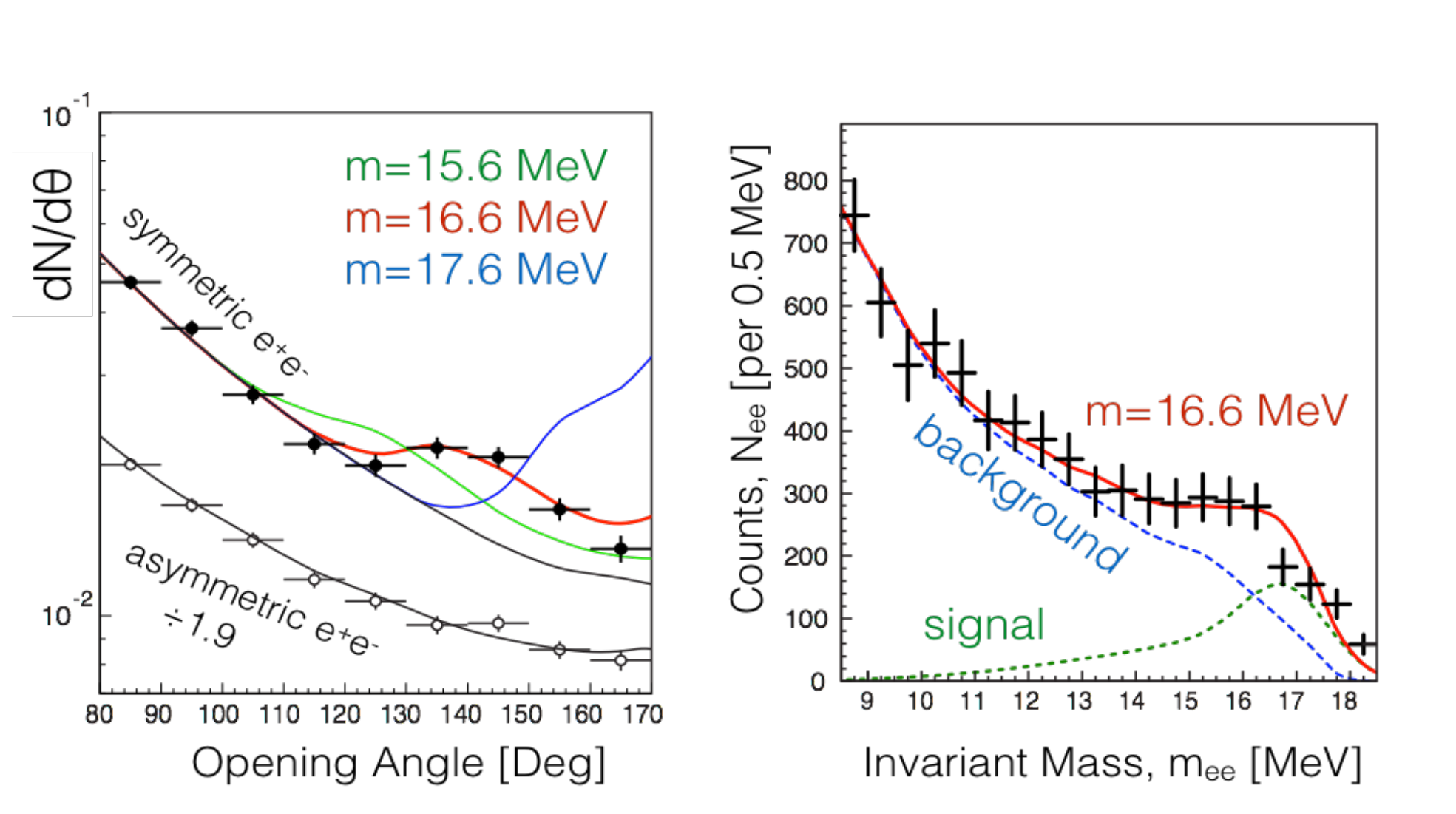}{\vspace{-4mm}}
  \caption{\small{Angular and invariant mass distribution of the internal conversion electron-positron pairs measured by the Atomki spectrometer (plots taken from  \cite{Krasznahorkay:2015iga}, modified by P. Tanedo).\vspace{1mm}}}
  \label{fig:6} 
\end{figure}

Instead of a monotonically decreasing Standard Model background of internal conversion  electron-positron pair angular distribution, the experiment shows a pronounced bump at the angle $\theta \approx  140^{\circ}$, which corresponds to an invariant mass:
\bea
m_{e^+e^-} = 16.7 \pm 0.35 \,(\rm stat) \pm 0.5 \,(\rm sys) \ {\rm MeV} \ ,
\eea
as shown on the plots in Fig.~\ref{fig:6}. The best fit to data was obtained for a new particle interpretation, in which case the significance of the signal is 6.8 sigma and the quality of the fit is excellent $(\chi^2/{\rm dof} = 1.07)$\footnote{No significant excess has been reported in the predominantly isovector transition from  $^8{\rm Be}(17.64)$ to the ground state in the original experiment of \cite{Krasznahorkay:2015iga}. However, recent preliminary claims in \cite{Krasznahorkay:2017qfd,Krasznahorkay:2017gwn} describe an excess in this transition compatible with a new particle interpretation.}.\vspace{2mm}

There are several properties of the signal adding credibility to the anomaly:  
\begin{itemize}
\item[$\rightarrow$] Due to the large number of recorded events it clearly is not  a statistical fluctuation.\vspace{0.0mm}
\item[$\rightarrow$] Signal is a bump,  not a ``last bin'' effect.\vspace{0.0mm}
\item[$\rightarrow$] It rises and falls when scanning through proton energies around the resonance.\vspace{0.0mm}
\item[$\rightarrow$] Only events passing the 18 MeV energy gate were recorded.\vspace{0.0mm}
\item[$\rightarrow$] Excess appears only for symmetric electron-positron pairs, which is expected in case of an intermediate massive particle. \vspace{0.0mm}
\item[$\rightarrow$] Peaks in angular and invariant mass distributions match.\vspace{2.0mm}
\end{itemize}

Obviously, an independent check of the result is necessary and all options have to be considered and thoroughly investigated: 
\begin{itemize}
\item[$\textcircled{\raisebox{-0.5pt}{\footnotesize{1}}}$]  Uncontrolled systematic errors, problem with the equipment or data recording. \vspace{-2.0mm}
\item[$\textcircled{\raisebox{-0.5pt}{\footnotesize{2}}}$]  Nuclear interference effects -- this possibility was actually explored by two theoretical groups \cite{Zhang:2017zap,Ward}. However, none of them provided an alternative explanation for the $^8{\rm Be}$ anomaly, with only \cite{Zhang:2017zap} revealing nuclear physics backgrounds that can only weaken the significance of the Atomki signal. \vspace{-2.0mm}
\item[$\textcircled{\raisebox{-0.5pt}{\footnotesize{3}}}$]   Finally, it can be an actual  sign of new physics, which is the option we concentrate on throughout the remainder of this review.\vspace{2mm}
\end{itemize}

The questions we focus on answering now are the following: What kind of particle can produce such a signal? What are its required couplings to Standard Model particles? Is this scenario consistent with all available data? Are there UV complete models accommodating the anomaly? Which other experiments can probe this?

\section{Particle Candidates}

The fact that  the signal consists of electron-positron pairs clearly indicates that if there is an intermediate particle decaying to them, it has to be a boson. Confining ourselves to the simplest options, we consider only particles with spin zero or one.

\subsection{Scalar}
The scalar case can be eliminated by symmetry considerations. The transition occurs from a $J^P \!=\! 1^+$ to a $0^+$ state. Angular momentum conservation requires the scalar to have $L\!=\!1$. On the other hand, parity conservation forces  $L$ to be even,  in contradiction to the first requirement. 

\subsection{Pseudoscalar}
An axion-like particle  generically couples to two photons through loop diagrams. 
For a mass of 17 MeV such a particle, in the absence of tree-level couplings to the Standard Model, is ruled out for the entire range of the effective pseudoscalar-photon-photon coupling \cite{Hewett:2012ns,Dobrich:2015jyk}. However, as argued in  \cite{Ellwanger:2016wfe}, one can overcome this constraint by postulating tree-level axial couplings of the pseudoscalar to quarks and leptons. In particular, a nuclear shell model calculation reveals \cite{Ellwanger:2016wfe} that a 17 MeV axion-like particle is capable of producing a signal similar to the one observed by the Atomki experiment, and, at the same time, remain consistent with  all other measurements. To accomplish this, a fine-tuning of the model parameters  is needed to suppress flavor changing neutral currents. 
The drawback  is that the required  couplings  do not fit into  a UV complete model. 

\subsection{Vector}
The interpretation of the $^8{\rm Be}$ anomaly as a sign of a new vector boson was exhaustively analyzed from a theoretical perspective in \cite{Feng:2016jff,Feng:2016ysn}. Experimental constraints require the coupling of the new vector boson to protons to be suppressed. This is the reason why the new boson was dubbed as ``protophobic''. 
Recent preliminary experimental results \cite{Krasznahorkay:2017qfd,Krasznahorkay:2017gwn} seem to favor the 17 MeV protophobic vector boson interpretation of the $^8{\rm Be}$ anomaly.
We discuss the details of this scenario in Sec.~\ref{protophobic}. 

\subsection{Pseudovector}

Suppression of the coupling to protons is not required in the axial vector case. 
Such a  pseudovector  was mentioned in  \cite{Feng:2016ysn} as a valid candidate, but the idea was not pursued due to lack of numerical results on $^8{\rm Be}$ nuclear transition matrix elements. A detailed numerical analysis was performed  only in  \cite{Kozaczuk:2016nma}, where the relevant matrix elements were calculated using \emph{ab initio} methods. It was shown that there exists a region in parameter space of the axial vector couplings to quarks and leptons  which successfully explains the $^8{\rm Be}$ anomaly and remains consistent with all other experiments \cite{Kozaczuk:2016nma}\footnote{Ref.~\cite{Dror:2017nsg} claims that meson decay bounds are still problematic for the model.}. It is also argued  that the axial vector candidate fits well into a UV complete model constructed in  \cite{Kahn:2016vjr}.

\subsection{Dark $Z$}
It is interesting to point out that a vector boson candidate with no definite parity and experiencing mixing with the Standard Model $Z$ boson cannot explain the $^8{\rm Be}$ anomaly due to extremely stringent bounds from atomic parity violation experiments \cite{Davoudiasl:2012qa}.

\begin{figure}[t!]
  \centering
      \includegraphics[width=0.4\textwidth]{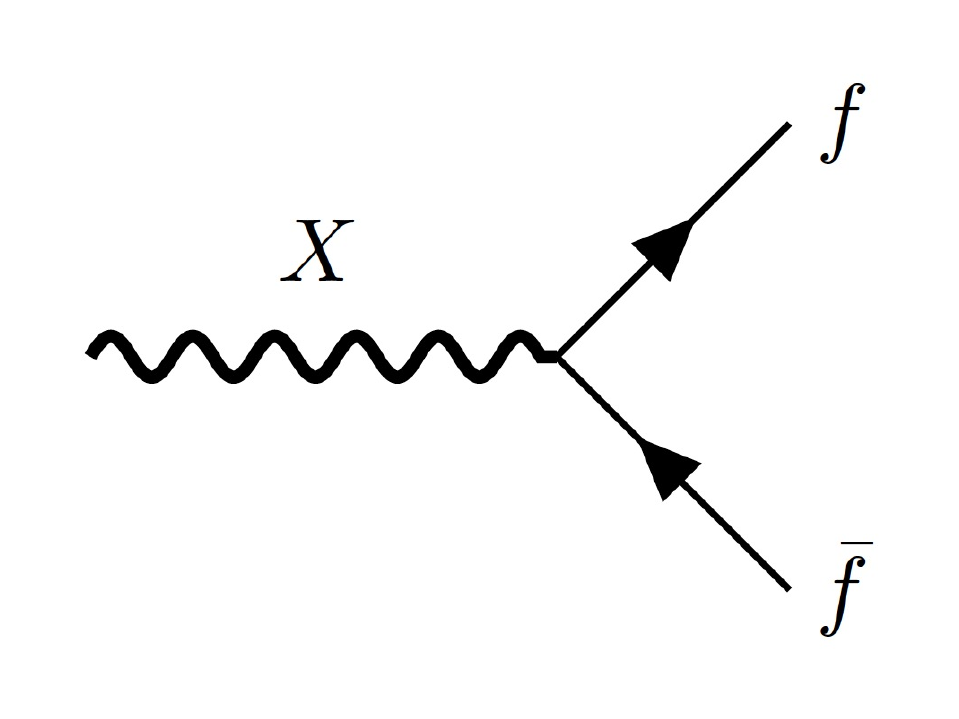}\vspace{-5mm}
  \caption{\small{Vector boson $X$ coupling to a fermion-antifermion pair.}}
  \label{fig:7}
\end{figure}

\section{Protophobic Boson}
\label{protophobic}

Let us denote the new vector boson by $X$ and write its coupling to fermions through the standard gauge and Lorentz invariant Lagrangian terms:
\bea
\mathcal{L} \supset - \,e \ \!X_\mu \sum_f \varepsilon_f \bar{f} \gamma^\mu f \ ,
\eea
corresponding to the Feynman diagram in Fig.~\ref{fig:7}.
In this notation $\varepsilon_f$ is the coupling strength of $X$ to a given fermion-antifermion pair in units of the electron charge.

\subsection{Coupling to quarks}
We discuss below  the signal requirements and the experimental constraints on the couplings of $X$ to the up and down quarks, arguing that $X$ cannot be a dark photon and explaining the origin of ``protophobia''.

\subsubsection{Signal strength}
Working in the effective theory regime in which the $^8{\rm Be}$ nucleus is treated as a fundamental degree of freedom, it can be shown that \cite{Feng:2016jff}:
\bea\label{Gamma}
\Gamma(^8{\rm Be}^* \rightarrow \,^8{\rm Be} \,X) = \frac{3e^2 (\varepsilon_u+\varepsilon_d)^2\, |\vec{p}_X|^3}{4\pi \Lambda^2} \big|\langle^8{\rm Be}|\,\bar{p} \gamma^\mu p + \bar{n} \gamma^\mu n\,|^8{\rm Be^*}\rangle\big|^2 \ ,
\eea
where $\Lambda \approx 10 \ {\rm MeV}$ is the energy scale up to which the effective theory is applicable. 
Because of the vector nature of $X$, one does not need to calculate the matrix elements in Eq.~(\ref{Gamma}), since they cancel against identical matrix elements for an electromagnetic transition involving a photon if we take the ratio of the two decay rates. The experimental value for this ratio is \cite{Krasznahorkay:2015iga}:
\bea\label{ratioGam}
\frac{\Gamma(^8{\rm Be}^* \rightarrow \, ^8{\rm Be} \,X)}{\Gamma(^8{\rm Be}^* \rightarrow \, ^8{\rm Be}\, \gamma)} = (\varepsilon_p+\varepsilon_n)^2 \frac{|\vec{p}_X|^3}{|\vec{p}_\gamma|^3} \approx 5.8 \times 10^{-6} \ ,
\eea
which implies
\bea\label{uplusd}
|\varepsilon_u+\varepsilon_d| \approx 3.7 \times 10^{-3} \ .
\eea
Therefore, the strength of the Atomki signal provides a constraint only on the sum of the $X$ couplings to the up and down quarks, and does not depend on the couplings to leptons. 

\subsubsection{Dark photon?}
It is important to point out that the new  boson cannot be a dark photon \cite{Kobzarev:1966qya,Okun:1982xi,Holdom:1985ag,Holdom:1986eq}, i.e., it cannot have couplings proportional to electric charge. The condition in Eq.~(\ref{uplusd}) would then give $\varepsilon \approx 0.01$, but this value is excluded by several experiments, including NA48/2 \cite{Batley:2015lha,Raggi:2015noa}. 

\subsubsection{Origin of ``protophobia''} 
\label{protophobica}
\begin{figure}[t!]
  \centering
      \includegraphics[width=0.35\textwidth]{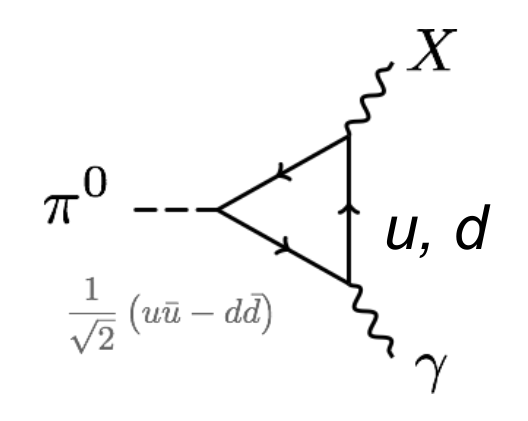}\vspace{-2mm}
  \caption{\small{Neutral pion decay channel searched for by the NA48/2 experiment. Null results provide constraints leading to the protophobic nature of $X$.}}
  \label{fig:8}
\end{figure}

Apart from the strength of the signal, the most relevant constraint on the couplings to quarks comes from the already mentioned NA48/2 experiment, which performs a search for new gauge bosons in the decays of neutral pions, as shown in Fig.~\ref{fig:8}.  
Those constraints affect a different combination of couplings to quarks, namely,
\bea\label{6}
|2\varepsilon_u +\varepsilon_d| < 8 \times 10^{-4} \ .
\eea
Combining Eq.~(\ref{6}) with the signal strength requirement in Eq.~(\ref{uplusd}) leads to 
\bea
-2.3 < \frac{\varepsilon_d}{\varepsilon_u}< -1.8 \ ,
\eea
which is equivalent to \cite{Feng:2016jff}
\bea
\label{forces}
-0.067< \frac{\varepsilon_p}{\varepsilon_n}< 0.078 \ .
\eea
The condition in Eq.~(\ref{forces}) forces $X$ to have a coupling to protons suppressed with respect to its coupling to neutrons (and individual quarks as well). This is the reason why the new gauge boson is called ``protophobic'' and the condition
\bea
\varepsilon_p=0
\eea
was adopted. This assumption removes the sensitivity of the model to future NA48/2 exclusion limits. 
\begin{figure}[t!]
  \centering
      \includegraphics[width=0.6\textwidth]{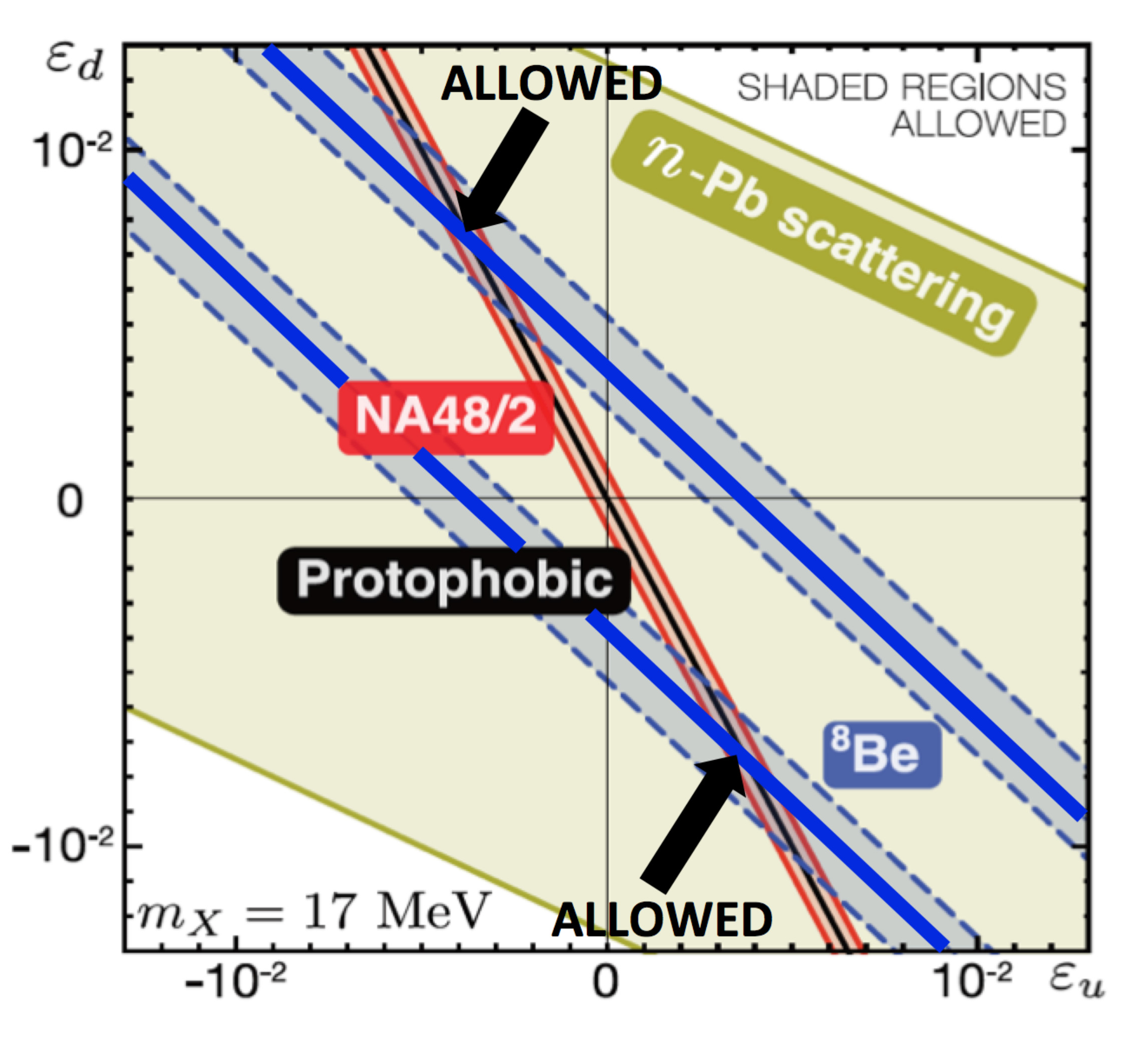}\vspace{-2mm}
  \caption{\small{Constraints on the protophobic vector boson $X$ coupling to quarks plotted in the $(\varepsilon_u, \varepsilon_d)$ plane. The only two tiny  allowed regions are indicated by arrows. Figure adopted from \cite{Feng:2016ysn} and modified.}}
  \label{fig:9}
\end{figure}
A summary of constraints on the couplings to quarks is presented graphically in Fig.~\ref{fig:9}.  For details regarding the less stringent constraint coming from $n\!-\!{\rm Pb}$ scattering \cite{Barbieri:1975xy} see \cite{Feng:2016ysn}.

\subsubsection{Isospin breaking effects}
\label{isospinmixing2}
As mentioned in footnote~\ref{foot1}, the analysis in Sec.~\ref{protophobica} was done with the simplifying assumption that the state $^8{\rm Be}(18.15)$ is a pure isospin singlet state, which is not exactly the case \cite{Pastore:2014oda,Wiringa:2000gb,Pieper:2004qw,Wiringa:2013fia}. The isospin eigenstate is actually a linear combination of the states $^8{\rm Be}(18.15)$ and $^8{\rm Be}(17.64)$. In particular, the higher energy isospin eigenstate can be written as \cite{Pastore:2014oda}:
\bea\label{isospinmixing}
|\Psi_T\rangle = \alpha\, |^8{\rm Be}(18.15)\rangle + \beta\,|^8{\rm Be}(17.64)\rangle \ ,
\eea
where $\alpha \simeq 0.98$ and $\beta \simeq 0.21$. 

\begin{figure}[t!]
  \centering
      \includegraphics[width=1\textwidth]{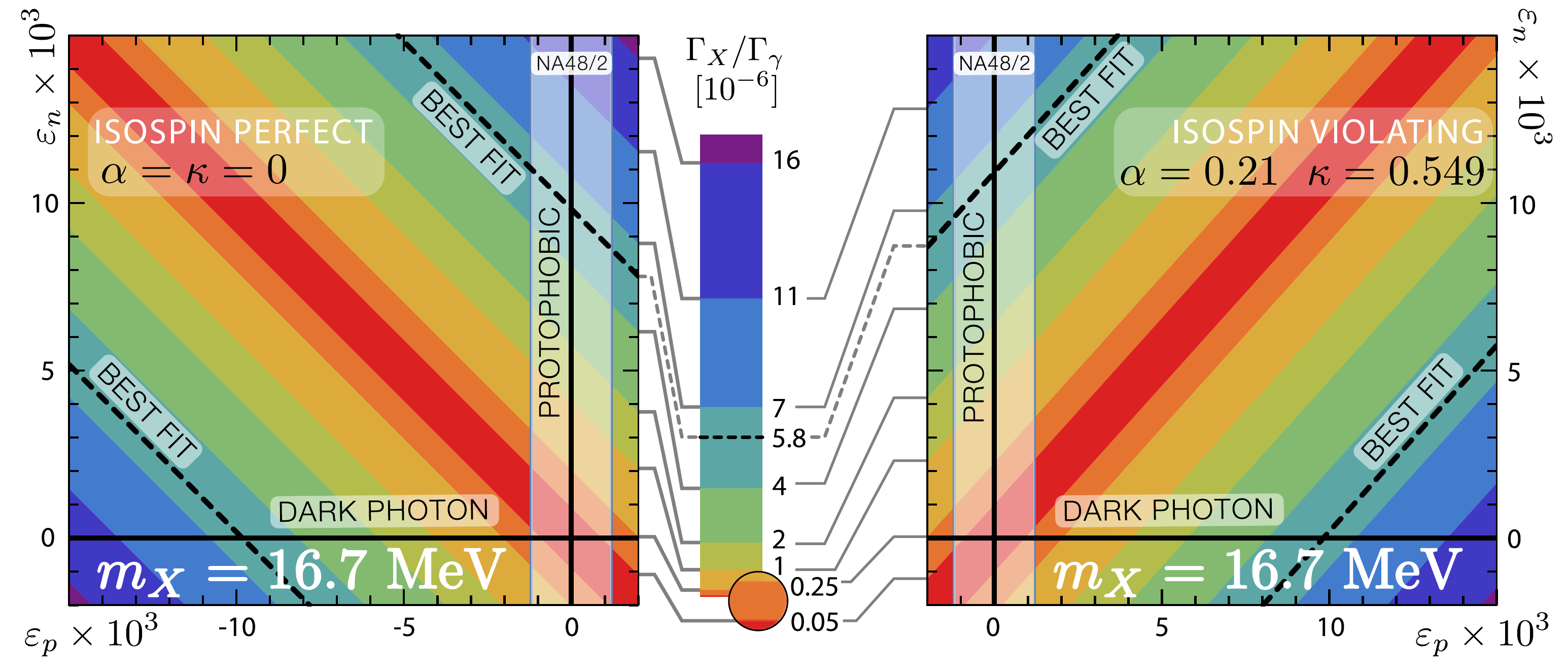}\vspace{-2mm}
  \caption{\small{Comparison of the ratio of decay rates in Eq.~(\ref{ratioGam}) without isospin violation effects taken into account (left plot) and the corresponding ratio in Eq.~(\ref{ratio2}) with the isospin mixing and breaking included (right plot). Figure adopted from  \cite{Feng:2016ysn}.}}
  \label{fig:10}
\end{figure}

Apart from isospin mixing, described by Eq.~(\ref{isospinmixing}), one should also take into account  isospin breaking effects \cite{Feng:2016ysn}. Those are measured by a parameter $\kappa$ and the best fit to spectroscopy data is obtained for $\kappa \approx 0.55$. For those values of parameters describing isospin mixing and breaking effects, the expression corresponding to Eq.~(\ref{ratioGam}) takes the form \cite{Feng:2016ysn}:
\bea\label{ratio2}
\frac{\Gamma(^8{\rm Be}^* \rightarrow \, ^8{\rm Be} \,X)}{\Gamma(^8{\rm Be}^* \rightarrow \, ^8{\rm Be}\, \gamma)} \simeq \big| \, 0.05\,(\varepsilon_p+\varepsilon_n) + 0.95\, (\varepsilon_p-\varepsilon_n) \big|^2 \frac{|\vec{p}_X|^3}{|\vec{p}_\gamma|^3}  \ .
\eea
Figure~\ref{fig:10} shows how the best fit value for the ratio of the Atomki signal to the photon background changes when isospin violation effects are taken into account. In general, those effects are significant. Surprisingly though, in the limit of pure protophobia, i.e., $\varepsilon_p = 0$, isospin violation introduces only a 20\% modification  to the isospin perfect case. 

\subsection{Coupling to leptons}
\begin{figure}[t!]
  \centering
      \includegraphics[width=0.55\textwidth]{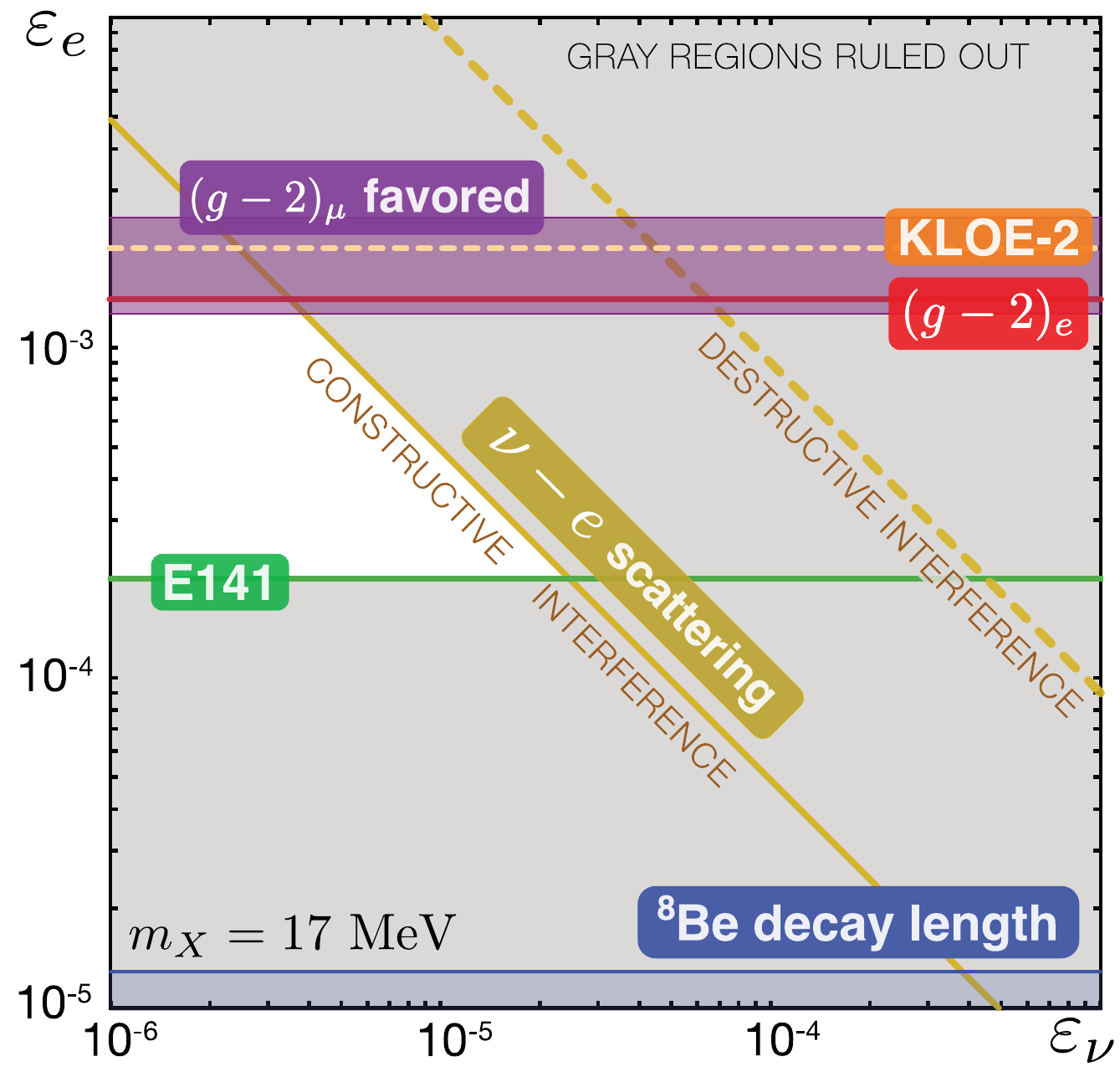}\vspace{-2mm}
  \caption{\small{Various constraints on the couplings of $X$ to leptons in the $(\varepsilon_\nu, \varepsilon_e)$ plane. If $\varepsilon_e \varepsilon_\nu > 0$, then only the white region is allowed. Figure adopted from  \cite{Feng:2016ysn}.}}
  \label{fig:11}
\end{figure}

Apart from constraints on the couplings of $X$ to quarks, there are also experimental limits on the allowed couplings to electrons and neutrinos. 
The obvious requirement is that $X$ interacts with electrons with a large enough coupling to decay within the detector. However, more stringent constraints than this come from experiments dedicated to new physics searches. The strongest lower limit on the $X$ coupling to electrons comes from the beam dump experiment SLAC E141 \cite{Riordan:1987aw,Bjorken:2009mm,Essig:2013lka}:
\bea
2\times 10^{-4} \lesssim |\varepsilon_e| \ ,
\eea
whereas the strongest upper limit is set by the electron $g\!-\!2$ searches \cite{Davoudiasl:2014kua}:
\bea
|\varepsilon_e|\lesssim 1.4 \times 10^{-3} \ .
\eea
Regarding the coupling to neutrinos, the most stringent bounds come from the TEXONO experiment \cite{Bilmis:2015lja,Khan:2016uon} and yield:
\bea
\begin{aligned}
\sqrt{\varepsilon_e \varepsilon_\nu} &< 7\times 10^{-5} \ \ \ \ \ \ {\rm for} \ \ \ \ \ \ \varepsilon_e \varepsilon_\nu > 0 \ ,\\
\sqrt{|\varepsilon_e \varepsilon_\nu|} &< 3\times 10^{-4} \ \ \ \ \ \ {\rm for} \ \ \ \ \ \ \varepsilon_e \varepsilon_\nu < 0 \ .
\end{aligned}
\eea
Figure~\ref{fig:11} summarizes the constraints presented above and shows also other less stringent limits discussed  in  \cite{Feng:2016ysn}.

\subsection{Summary of constraints}

Collecting all the experimental bounds, the couplings of the 17 MeV protophobic $X$ boson to the Standard Model first generation fermions have to fulfill the following conditions:
\bea\label{summary}
\begin{aligned}
\varepsilon_u  &\approx \pm \,3.7 \times 10^{-3} \ ,\\
\varepsilon_d  &\approx \mp \,7.4 \times 10^{-3}\ ,\\
2\times10^{-4} \lesssim |\varepsilon_e| &\lesssim 1.4 \times 10^{-3}\ , \hspace{20mm}\\
\sqrt{\varepsilon_e\varepsilon_\nu} &\lesssim 7\times 10^{-5} \ ,
\end{aligned}
\eea
where the conservative  limit on $\varepsilon_\nu$ was adopted. Such milli-charged couplings to neutrons and electrons, along with a suppressed coupling to protons and neutrinos, are quite challenging to obtain from  a UV complete model. In Sec.~\ref{UV} we briefly review the few high energy completions for the protophobic boson constructed so far.

Finally, let us note that a protophobic boson can produce an excess in the internal conversion electron-positron pair angular distribution not only for the predominantly isoscalar transition from $^8{\rm Be}(18.15)$ to the ground state, but also for the predominantly isovector transition from $^8{\rm Be}(17.64)$ to the ground state. As argued in  \cite{Feng:2016jff}, the protophobic gauge boson contribution to such a transition is sufficiently kinematically suppressed not to be seen in the original Atomki experiment. However, as predicted in  \cite{Feng:2016ysn}, it should become visible with more accumulated data. It is very promising for the protophobic boson interpretation that the preliminary results obtained with more data \cite{Krasznahorkay:2017qfd,Krasznahorkay:2017gwn} indicate an excess also for the second transition.

\section{UV Complete Models}
\label{UV}
One may wonder -- just how natural, from a theoretical point of view, is the existence  of a 17 MeV  protophobic boson at low energies with the couplings to Standard Model particles summarized in Eq.~(\ref{summary}).
The only fundamental vector particles we have discovered so far are gauge bosons corresponding  to local symmetries. This hints that one can try to include the new protophobic boson into this picture precisely by associating it with a new  gauge group. The simplest option is to introduce an extra local  ${\rm U(1)}$  symmetry. 

One possible model-building path is to relate the additional gauge group to the global symmetries already existing in the Standard Model, in particular baryon number, lepton number, and, after electroweak symmetry breaking, electric charge. This  path was chosen in  \cite{Feng:2016ysn} and two models were proposed, one based on gauged ${\rm U(1)}_B$ and the second based on gauged ${\rm U(1)}_{B-L}$. In both cases after electroweak symmetry breaking  a kinetic mixing between the new gauge $\rm U(1)$ and $\rm U(1)_{\rm EM}$ is generated, and the resulting gauge boson couples to $B\!-\!Q$ and $B\!-\!L\!-\!Q$, respectively, precisely reproducing a protophobic scenario. 
 
The other path is to introduce a ``dark'' $\rm U(1)$ gauge group without relating it to existing global symmetries and choose the charges of the Standard Model particles under this gauge group to produce ``protophobia''. This method, however, requires the Standard Model Higgs to be charged under the new gauge group, which produces dangerous mass mixing between the Standard Model gauge boson $Z$ and the protophobic gauge boson $X$. This mixing is tightly constrained by atomic parity violation experiments \cite{Davoudiasl:2012qa}.

\subsection{Gauged baryon number}
It has recently been shown that experimentally viable models with gauged baryon number can be constructed \cite{Duerr:2013dza,Arnold:2013qja,Duerr:2014wra,Perez:2014qfa,Fornal:2015boa}. Such theories contain a new gauge boson which couples to baryon number and, in their simplest realization, require six extra fermionic fields to cancel the arising gauge anomalies. The lightest neutral field among them  is a good dark matter candidate.
It turns out that the model constructed in  \cite{Duerr:2013dza} can be used as a UV completion for the protophobic gauge boson \cite{Feng:2016ysn}.
Denoting the new gauge coupling (in units of electric charge) by $\varepsilon_B$ and the coefficient of the term mixing $\rm U(1)_B$ and $\rm U(1)_{\rm EM}$ after electroweak breaking by $\varepsilon$, the coupling of fermions to the $X$ gauge boson is given by:
\bea
\varepsilon_f = \varepsilon_B B_f + \varepsilon\, Q_f \ ,
\eea
where $B_f$ is the fermion's baryon number and $Q_f$ is its electric charge. 
It is immediately evident that there is no coupling between $X$ and the neutrinos, which automatically fulfills the neutrino constraint in Eq.~(\ref{summary}). 
Furthermore, it is possible to find values of $\varepsilon_B$ and $\varepsilon$ which produce the couplings to quarks and electrons fulfilling the remaining conditions in Eq.~(\ref{summary}). 
The mass of the protophobic gauge boson in this model is given by:
\bea
m_X = 3\,e\, |\varepsilon_B|\, v_B \ ,
\eea
where $v_B$ is the vacuum expectation value of the Higgs field breaking $\rm U(1)_B$. With $m_X \!\approx\! 17 \ {\rm MeV}$ one requires $v_B \!\approx\! 10 \ {\rm GeV}$, which is a very low  
${\rm U}(1)_B$ breaking scale \cite{Farzan:2016wym}. As a result, the new fermions needed to cancel gauge anomalies cannot develop large vector-like masses and are quite tightly constrained by LHC searches for new particles at the electroweak scale. The other source of constraints comes from electroweak precision measurements and the Standard Model Higgs decays. However, as shown in  \cite{Feng:2016ysn}, there exists a region in parameter space where the gauged $\rm U(1)_B$ model for the 17 MeV protophobic gauge boson is allowed. We note that  \cite{Dror:2017nsg,Dror:2017ehi} claim the meson decay bounds are still problematic for the model and require further modifications. The applicability of those bounds along with possible resolutions are currently under investigation \cite{Tim}. Finally, let us mention that assuming  $\varepsilon_\mu \approx \varepsilon_e$  removes,
at least partially, the long-standing discrepancy in $(g-2)_\mu$ between measurements \cite{Bennett:2006fi} and the Standard Model prediction \cite{Hagiwara:2006jt}.

\subsection{Gauged $B\!-\!L$}
Perhaps a more appealing  model is the one based on the gauge group ${\rm U(1)}_{B-L}$ \cite{Mohapatra:1980qe,Buchmuller:1991ce,Buchmuller:1992qc}, which can  also be used to construct a UV complete theory of the protophobic gauge boson \cite{Feng:2016ysn,Seto:2016pks}. It has the nice property of being anomaly-free already upon including right-handed neutrinos, which itself allows for the seesaw mechanism \cite{Minkowski:1977sc}. A gauged ${\rm U(1)}_{B-L}$ is also present in the symmetry breaking pattern of the $\rm SO(10)$ grand unified theory \cite{Fritzsch:1974nn} leading to the Standard Model. 

Denoting the new gauge coupling (in units of electric charge) by $\varepsilon_{B-L}$,  the interaction strength  between fermions and the $X$ gauge boson is given by:
\bea
\varepsilon_f = \varepsilon_{B-L} (B-L)_f + \varepsilon\, Q_f \ .
\eea
The complication now is that  an unsuppressed coupling to  neutrinos is initially present. However, the neutrino $X$-charge can be neutralized  by introducing new vector-like leptons and mixing them with the active neutrinos. As in the previous case, the model requires a low  ${\rm U}(1)_{B-L}$ breaking scale \cite{Bilmis:2015lja,Carlson:1986cu,Heeck:2014zfa,Lee:2016ief}, but it is shown to circumvent all experimental bounds \cite{Feng:2016ysn}, while providing a range of parameters consistent with the $^8{\rm Be}$ anomaly. As in the previous case, \cite{Dror:2017nsg} argues that the model requires further modifications  to eliminate problematic meson decay constraints. A more careful investigation of this is currently being carried out \cite{Tim}.

\subsection{Other models}
\label{otherm}
An alternative model-building strategy is to simply assign ``dark'' charges to the Standard Model quarks and leptons, such that the dark gauge boson does not couple to protons
and fulfills all other requirements in Eq.~(\ref{summary}). This approach has been pursued by several authors \cite{Gu:2016ege,DelleRose:2017xil}. The problem is that in such scenarios there  exists a generic mass mixing between the new gauge boson and the Standard Model $Z$, severely constrained by atomic parity violation experiments \cite{Davoudiasl:2012qa}. This can be fixed, for example, by extending the gauge group by yet another $U(1)$  and introducing sufficiently large mixing between them \cite{Gu:2016ege,DelleRose:2017xil,Chen:2016tdz}. 

Several other models focused on the dark matter sector were constructed with the protophobic gauge boson mediating interactions between the dark sector and the Standard Model \cite{Jia:2016uxs,Kitahara:2016zyb}. It was shown that the correct dark matter relic density can be obtained for the parameter region relevant for the $^8{\rm Be}$ anomaly. 

Finally, one can also construct the new gauge group using just the Standard Model symmetries (based on the charges $Q$, $B$, $L$) other than ${\rm U}(1)_B$ or ${\rm U}(1)_{B-L}$ \cite{Chen:2016tdz,Fayet:2016nyc} with sufficient kinetic mixing to explain the Atomki signal.

\vspace{3mm}

\section{Experimental Verification}

\begin{figure}[t!]
  \centering
      \includegraphics[width=0.65\textwidth]{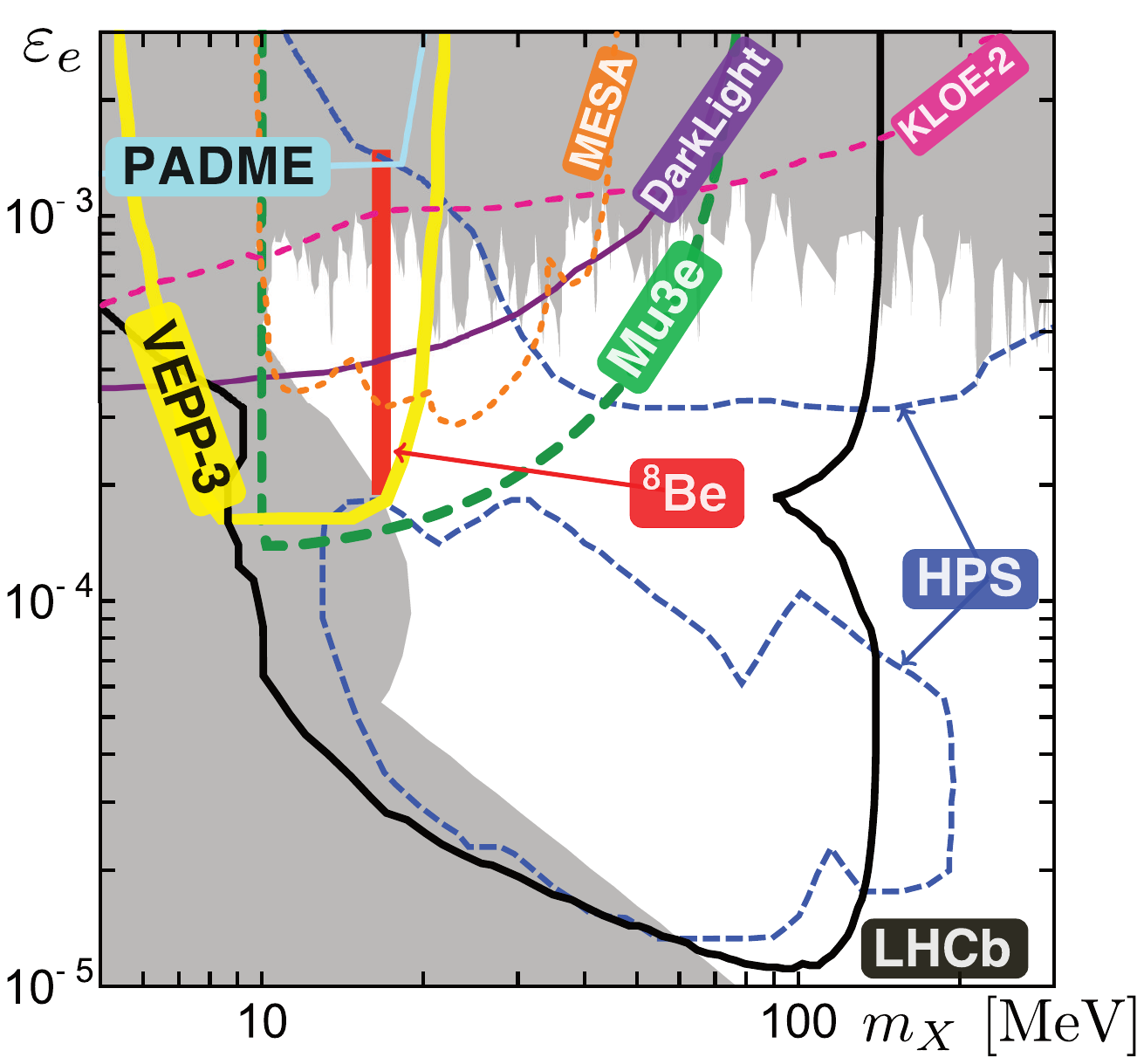}
  \caption{\small{Future sensitivity of current and upcoming experiments to the parameter range in the $(m_X,\varepsilon_e)$ plane relevant for the $^8{\rm Be}$ anomaly (with the predicted timeline described in the text). Figure adopted from  \cite{Feng:2016ysn}.}\vspace{4mm}}
  \label{fig:12}
\end{figure}

Further experiments are crucial in determining the real nature of the $^8{\rm Be}$ anomaly, whether it will be confirming or refuting the new physics interpretation. However, to fully test the anomaly a different experimental setup should be used. 
One option might be to look at other sufficiently energetic {\bf nuclear electromagnetic transitions}. As mentioned in Sec.~\ref{newphysics}, some of the very limited options are: $^{10}{\rm Be}(17.8)$, $^{10}{\rm B}(18.4)$ and $^{10}{\rm B}(19.3)$ \cite{Ling:1971oik,Subotic:1978mab}.
Other ongoing  and future experiments relevant for probing the protophobic gauge boson are shown in Fig.~\ref{fig:12} along with their reach in parameter space.

A definitive answer will be provided by the {\bf LHCb} experiment \cite{Ilten:2015hya} (black line) after analyzing data from Run 3 scheduled for the years 2021-2023. It will probe the entire parameter region relevant for the anomaly by looking at  charmed meson decays involving a dark photon. The {\bf Mu3e} experiment \cite{Echenard:2014lma} (green line) is also capable of scanning through the whole region of interest during its second phase starting in 2018. It is designed to look for anomalous $\mu^+$ decays involving a dark photon. The only other experiment probing the entire $^8{\rm Be}$ region will be {\bf VEPP-3} \cite{Wojtsekhowski:2012zq,Rachek:2017gdc}, a proposal of which has been submitted, and, once accepted, will take three to four years. It will look for missing mass spectra in a positron beam interacting with electrons in a gaseous hydrogen target.   

Partial information regarding the $^8{\rm Be}$ parameter space will be supplied by several other experiments.  {\bf KLOE-2} \cite{Anastasi:2015qla,delRio:2016anz} (pink line) will look for dark photon signatures in electron-positron collisions. The sensitivity plotted in Fig.~\ref{fig:12} should be reached in 2018. The commissioning of the {\bf MESA}  experiment \cite{Beranek:2013yqa} (orange line), looking for dark photons in positrons colliding with electrons in a gaseous target, is scheduled for the year 2020. Finally, {\bf DarkLight} \cite{Balewski:2014pxa} (purple line) will also look for dark photons in positron-electron interactions, with the plotted sensitivity reached in two to three years. Other experiments, like {HPS} \cite{Moreno:2013mja} and  {PADME}  \cite{Raggi:2014zpa,Raggi:2015gza,Chiodini:2017fer} are barely sensitive to the $^8{\rm Be}$ region of interest. 

Apart from the above-mentioned efforts, other methods might also be useful in searching for the protophobic gauge boson. Those include isotope shift spectroscopy  \cite{Frugiuele:2016rii,Berengut:2017zuo}  and  kaon decays  \cite{Chen:2016kxw,Chiang:2016cyf,Pospelov:2017pbt}.
All in all, within ten years from now the fate of the protophobic gauge boson should be revealed.

\section{Summary}

There exists an interesting anomaly in beryllium-8 nuclear transitions discovered by a group led by A.~Krasznahorkay at the Atomki Institute for Nuclear Research in Debrecen, Hungary.
The signal is a bump in the angular and invariant mass distributions of internal conversion electron-positron pairs. It has been shown that the excess can be explained by a new 17 MeV boson, with a  significance of almost seven sigma.

Several new physics interpretations of the anomaly have been proposed in the literature. Those include a new vector boson, an axial vector, and a pseudoscalar. In this review we focused primarily  on the first option and showed that experimental constraints force such a vector boson to have suppressed coupling to protons, hence the name protophobic. We also discussed UV complete models that result in a 17 MeV protophobic gauge boson at low energies, and are consistent with all available experimental data.

It is interesting that the upgraded Atomki experiment recorded a preliminary signal in another beryllium-8 nuclear transition compatible with the protophobic gauge boson interpretation. Of course further experimental efforts  are crucial in determining the true origin of the anomaly, and we enumerated other current and upcoming experiments which can be used to confirm or exclude the protophobic gauge boson interpretation.

\section*{Acknowledgments}
I am grateful to Jonathan Feng, Tim Tait, Susan Gardner, Philip Tanedo, Iftah Galon and Jordan Smolinsky for a fruitful collaboration. 
This research was supported in part by the DOE
grant \#DE-SC0009919 and the NSF grant \#PHY-1316792.

\bibliographystyle{utphys}
\bibliography{8Be_review}

\end{document}